\documentclass[twocolumn,prb]{revtex4}
\usepackage{graphicx}
\usepackage{bm}
\usepackage{amsmath}
\begin{document} \title{Theory of Spin Orientation of Semiconductor Carriers
at a Ferromagnetic Interface}

\author{J. P. McGuire, C. Ciuti, and L. J. Sham}
\affiliation{Department of Physics, University of California San Diego,
La Jolla CA 92093-0319.}

\begin{abstract}

A quantum theory of the spin-dependent scattering of semiconductor
electrons by a Schottky barrier at an interface with a ferromagnet
is presented.  The reflection of unpolarized non-equilibrium
carriers produces spontaneous spin-polarization in the
semiconductor.  If a net spin-polarization pre-exists in the
semiconductor, the combination of the ferromagnet magnetization
and the incident carrier polarization combine to tilt the
reflected polarization in the semiconductor. The spin reflection
properties are investigated as functions of the system
characteristics: the Schottky barrier height, semiconductor doping
and applied bias.  The effect on reflection due to the variation
of the barrier width with electron energy is contrasted for two
means of excitation: optical or electrical.  Optically excited
electrons have a wider energy spread than the near-equilibrium
excitation from non-magnetic ohmic contacts.

\end{abstract}

\date{\today}
\maketitle

\section{Introduction}

Semiconductor spintronics \cite{spinreview} is a new field in
which the manipulation of carrier spins is a central issue.  The
first device proposal in this field was by Datta and Das,
\cite{datta} which consisted of (1) spin injection from a
ferromagnet source contact into a semiconductor two-dimensional
electron gas (2DEG), (2) manipulation of the spin with a gate bias
through the Rashba spin-orbit effect, \cite{rashba} and (3)
measurement of the spin with a ferromagnetic drain contact.
Progress has recently been made  both in spin injection
\cite{fiederling,ohno,zhu,hanbicki,johnston} and in control via
the Rashba effect \cite{nitta} but it appears that there are
difficulties in practical implementation of this kind of device.
The demand of sufficiently strong spin-orbit interaction requires
a narrow gap semiconductor. More basically, the spin-orbit
interaction driven by the  electric field normal to the plane of
the electron gas polarizes the electron spin in the plane normal
to its wave vector. To reduce spin-cancellation effects among
different wave-vectors, \cite{comment} a one-dimensional channel
was suggested.  \cite{datta} In this paper, we describe in theory
a different approach to generate and control spin polarization in
semiconductors which is based on the spin-dependent properties of
thin semiconductor layers in close proximity to a ferromagnet.

Time-resolved Faraday rotation experiments over the last decade
have given much insight into electron spin dynamics in
semiconductors. \cite{faraday}  In particular, it has been shown
that in lightly doped semiconductors the electron spin coherence
can persist for hundreds of nanoseconds, \cite{kikkawa} a packet
of spin-coherent electrons can be dragged microns using an
electrical bias \cite{drag} and can maintain their coherence
through an interface between two semiconductors, \cite{malajovich}
and electron spin coherence can produce large nuclear
effects\cite{nmr} in the semiconductor through the hyperfine
interaction \cite{orientation}.  Recent pump-probe experiments
have been performed on $n$-doped semiconductor epilayers in
contact with a ferromagnet, demonstrating that the proximity of
the ferromagnet can induce large nuclear fields
\cite{fmimprinting} and spontaneous electron spin polarization
\cite{fmcoherence} in the semiconductor.  In a recent letter,
\cite{ciuti1} we gave a theory of spin-dependent reflection at the
semiconductor-ferromagnet Schottky barrier as an explanation of
the origin of the spin polarization. In this long paper, we
provide a detailed treatment of the spin reflection. We also
include calculated results of applying electric bias between the
semiconductor and the ferromagnet, stimulated by an ongoing
experiment under bias.\cite{epstein} We also investigate the
different spin reflection properties resulting from optical
pumping and from electrical excitation of non-equilibrium carriers
in the semiconductor.

The paper is organized as follows.  Section \ref{reflection} gives
a description of the physics of semiconductor electron spin
polarization upon reflection at a semiconductor-ferromagnet
interface. In Section \ref{matrix} we present the general
scattering theory.  In Section \ref{example}, we apply the theory
to the case of Schottky junction, showing the role of the Schottky
barrier, semiconductor doping and applied bias. Moreover, we
compare the case of optical excitation to that of electrical
excitation. Conclusions are drawn in Section \ref{conclusions}.

\section{Spin Reflection Off a Ferromagnet}
\label{reflection}

The generation of spin polarization by scattering of spin
particles against a spin-polarized target has a long
history.\cite{mott}  We consider the situation where a
non-equilibrium distribution of carriers is injected in the
semiconductor (either electrically or optically) and determine the
transient spin-dependent reflection dynamics.  The semiconductor
is assumed to be $n$-doped to ensure a reasonably thin Schottky
barrier (for sizeable quantum-mechanical coupling between the
semiconductor and the ferromagnet) and reduced spin relaxation.
Thus, an initially spin-compensated group of excited electrons
will be reflected by the ferromagnetic interface with a net spin
polarization.  Since the momentum relaxation of the
non-equilibrium carriers is much faster than their spin
relaxation, the Fermi sea in the semiconductor conduction band is
left with a spin polarization. In this way, the spin reflection
produces a ferromagnetic ``imprinting'' of the semiconductor
electrons. More generally, if the non-equilibrium electrons have a
pre-existing spin polarization, the ferromagnetic imprinting
manifests itself as a tilting of the original polarization vector.
Thus, the spin reflection not only generates a spin polarization,
but can also rotate a pre-existing ensemble spin.

In the case of optical excitation one should also consider the
spin dynamics of the excited holes in the valence band.  However,
we will neglect any spin effects from the valence band holes.  The
holes can gain polarization from the reflection at the interface
exactly as the conduction electrons, but the corresponding spin
polarization is known to decay very fast due to valence
band-mixing.\cite{uenoyama}  In particular the hole spin
relaxation time is much shorter than the optical electron-hole
recombination time.  In the case of optical excitation, a moderate
electron doping is essential to provide very long spin
lifetimes\cite{kikkawa}, because the Fermi sea acts as a spin
reservoir. The holes will recombine with electrons from the Fermi
sea and thus remove a strong source of electron relaxation via
exchange with the holes.\cite{bap}

\section{Theory of spin-dependent reflection}
\label{matrix}

To capture the essential physics of the problem we work with a
simplified effective mass Hamiltonian in both the semiconductor
and the ferromagnet.\cite{slon} The effective mass approximation,
though suitable for semiconductor heterostructures, can not
possibly account for all band properties of the ferromagnetic
metal.\cite{monier} We shall investigate one improvement of the
wave function matching between the dissimilar metal and
semiconductor media.\cite{einevoll}  The general results we
present should remain valid with more realistic calculations.

The Hamiltonian  of the metal/semiconductor junction is given by
\begin{eqnarray}
\nonumber H = &-& \frac{\hbar^2}{2} \frac{d}{dz} \left[
\frac{1}{m^\star(z)} \frac{d}{dz} \right] +
U(z) \\
& + & \frac{\Delta}{2} {\bm \sigma} \cdot \hat{\textbf M} ~ \Theta
(z) + \frac{g^{\star}}{2}  ~\mu_{\text B} {\bm \sigma} \cdot {\bf
B}_{\text{tot}} \Theta (-z) ~, \label{hamiltonian}
\end{eqnarray}
wher the $z$-axis is along the growth direction.  The
first term is the kinetic energy, where $m^\star (z)$ is the
 effective mass which is different for each region.  The second term
represents the spin-independent part of the potential energy. In
the semiconductor region ($z < 0$), the potential $U(z)$ produces
the band-bending due to the space charge layer associated with the
Schottky barrier.  $U(z)$ can be tailored by proper
heterostructure engineering (for example by inserting a
delta-doped layer at the interface) or by applying {\it in situ}
an electrostatic bias.  As we will show in detail in Section
\ref{example}, the profile of $U(z)$ can drastically modify the
spin-dependent coupling between semiconductor and ferromagnet. The
second line of Eq. (\ref{hamiltonian}) contains the spin-dependent
part of the Hamiltonian, the exchange interaction operator in the
ferromagnet ($z > 0$) and the Zeeman energy in the semiconductor
($z < 0$). $\Delta$ is the exchange splitting energy between the
majority and minority spin bands in the ferromagnet and
$\hat{\textbf M}$ is the unit vector along the direction of the
ferromagnet magnetization. In the Zeeman term, $g^\star$ is the
effective electron g-factor ($g^\star = -0.44$ for GaAs),
$\mu_{\text B}$ is the Bohr magneton, and the magnetic field ${\bf
B}_{\text{tot}}$ is the sum of the external field and the induced
nuclear field, ${\bf B}_{\text{tot}} = {\bf B} + {\bf
B}_{\text{N}}$. The Zeeman splitting has a negligible effect on
the spin polarization because it is typically several orders of
magnitude smaller than the exchange splitting in the ferromagnet
and the Schottky barrier height.  However, a weak magnetic field
is useful as a probe of the ferromagnetic imprinting, because it
induces a Larmor precession of the reflection-induced spin
polarization which can be detected, for example, by time-resolved
Faraday rotation.\cite{fmimprinting} The origin of the nuclear
field ${\bf B}_{\text{N}}$ is the hyperfine coupling between the
electron and the nuclear spins.  If the electron polarization has
a component along the external magnetic field vector, the electron
polarization is known to induce dynamically a nuclear spin
polarization through the Overhauser effect.\cite{fmimprinting} The
dynamically polarized nuclei produce an effective magnetic field
that acts back on the electrons
\begin{equation}
{\bf B}_{\text{N}} \sim \frac {g^\star}{|g^\star|} \frac{({\bf S} \cdot
{\bf B}){\bf B}}{B^2+B_0^2}~.
\end{equation}
Depending on the orientation of the electron spin polarization
vector ${\bf S}$ relative to ${\bf B}$, the nuclei can align
either along or against the external applied field. $B_0$ is a
phenomenological parameter to account for the fact that at low
applied field the nuclei are unable to align with the applied
field due to nuclear spin-spin interactions which tend to destroy
the dynamic nuclear polarization.

We denote a majority spin in the ferromagnet as $|+ \rangle$ and a
minority spin in the ferromagnet as $|- \rangle$, which are
eigenstates of the exchange splitting operator, ${\bm \sigma}
\cdot \hat{\textbf M} | \pm \rangle = \mp | \pm \rangle$ (the
magnetization is antiparallel to the net electron spin). In the
following, we will work explicitly in this basis.  In general, the
reflection of semiconductor electrons at the ferromagnetic
interface is represented by the reflection matrix $\hat{r}({\bf
k})$, which, in the ferromagnet spin basis, is
\begin{equation}
\hat{r}({\bf k}) =
\left (
\begin{array}{cc}
r_{+,{\bf k}} & 0 \\
0 & r_{-,{\bf k}} \\
\end{array}
\right )~,
\end{equation}
where $r_{+,{\bf k}}$ is the reflection coefficient for a
semiconductor electron with its spin aligned with the majority
spin band in the ferromagnet and $r_{-,{\bf k}}$ likewise with the
minority spin band.  This can be expressed in the vectorial form
\begin{equation}
\hat{r}({\bf k}) = \frac{1}{2} \left[ (r_{-,{\bf k}} + r_{+,{\bf
k}}) \openone +  (r_{-,{\bf k}} - r_{+,{\bf k}})  \hat{\bf M}
\cdot {\bm \sigma} \right] ,
\end{equation}
where $\openone$ is the unit matrix.

\begin{figure}
\includegraphics[width=6cm]{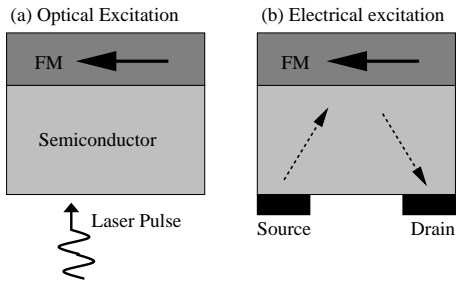}
\caption{The excitation of non-equilibrium electrons into a
semiconductor.  In (a), a short laser pulse with energy at or
above the bandgap of the semiconductor creates a distribution of
electrons related to the laser spectrum.  In (b), a source contact or an
STM tip
can inject electrons at or above the Fermi level in the semiconductor.  In
both cases the non-equilibrium electrons can reflect off the
interface with a ferromagnet. } \label{excitation}
\end{figure}
Suppose that a short excitation pulse injects non-equilibrium
electrons into the semiconductor. This perturbation can be applied
either by a short optical pump pulse or electrically through
lateral contacts or an STM tip (see Fig. \ref{excitation}).
Initially, the non-equilibrium electron spin density matrix has
the form
\begin{equation}
\hat{\rho}^{\rm i} ({\bf k},t=0)=\frac{1}{2} f^{\rm i}(k)
(\openone + {\bf P}^{\rm i} \cdot {\bm \sigma}),
\end{equation}
with $f^{\rm i}(k)$ the distribution of injected electrons (for
optical excitation it depends on the laser spectrum).  The initial
polarization is determined by the perturbation. Optically, the
polarization is determined by the usual selection rules, namely
${\bf P}^{\rm i} = 0$ for linearly polarized light and ${\bf
P}^{\rm i} \ne 0$ for elliptically polarized light.  In the case
of electrical excitation, an initial polarization can be created
through injection from magnetic contacts through ordinary spin
injection. Upon striking the interface, the non-equilibrium
electrons will reflect and transmit, and, to fully account for the
polarization in the semiconductor, we must calculate the effect of
this reflection on the spin density matrix:
\begin{eqnarray}
\nonumber \hat{\rho}^{\rm r}({\bf k},t) &=&  \hat{r}({\bf k})
\hat{\rho}^{\rm i}({\bf k},t) \hat{r}^{\dagger}({\bf k}) \\ &=&
f^{\rm i}(k,t) \frac{1}{2} [R_0({\bf k}) \openone + {\bf R}({\bf
k}) \cdot {\bm \sigma} ],
\end{eqnarray} where, with the ${\bf k}$-dependence understood,
\begin{eqnarray} R_0 &=& \frac{1}{2}[(|r_-|^2 + |r_+|^2) + (|r_-|^2 - |r_+|^2)
~\hat{\bf M} \cdot {\bf P}^{\rm i}] , \label{r0} \\ {\bf R} &=&
\frac{1}{2}[ (|r_-|^2 - |r_+|^2) + (|r_-|^2 + |r_+|^2)~ \hat{\bf
M} \cdot {\bf P}^{\rm i} ] ~\hat{\bf M} \nonumber \\ &+& {\rm
Re}(r_- r_+^*)~ (\hat{\bf M}\times {\bf P}^{\rm i} )\times
\hat{\bf M} - {\rm Im}(r_- r_+^*)~\hat{\bf M} \times {\bf P}^{\rm
i}~. \nonumber
\\ \label{polar}
\end{eqnarray}
In general the polarization after reflection ${\bf R}({\bf k})$ is
different from the original polarization ${\bf P}^{\rm i}$ of the
excited electrons in the semiconductor.

Since the electron distribution is not in equilibrium, the
relaxation of the spin density matrix will be dominated by the
relaxation of the hot carrier distribution,$f^{\rm i}(k,t) =
f^{\rm i}(k)exp{(-t/\tau_k})$. This relaxation is spin-independent
because it occurs on a much faster time scale than the
spin-relaxation time.  Thus, the reflection-induced
spin-polarization will leave a spin excitation in the
semiconductor electron sea. In order to quantitatively determine
the effect, we need to calculate the current flow into the
ferromagnet during the non-equilibrium transient,
\begin{equation}
\hat{j}(t) = \hat{j}^{\rm i}(t) + \hat{j}^{\rm r}(t) =
\int_{k_{\rm z} > 0} \frac{d^{3} {\bf k}}{(2 \pi)^3}
[\hat{\rho}^{\rm i}({\bf k},t) - \hat{\rho}^{\rm r}({\bf k},t)]
v_z ,
\end{equation} where $v_z =\hbar k_z/m^{\star}_{\text{sc}}> 0$ is the
velocity component normal to the interface.  This current flow
will be spin-dependent, so that the net spin in the semiconductor
from the reflection will be the negative of the spin transmitted
into the ferromagnet,
\begin{eqnarray} {\bf S^{\rm r}} &=& - {\rm Tr} \left\{\frac{\hbar}{2}{\bm
\sigma}
\int dt~ [\hat{j}^{\rm i}(t) + \hat{j}^{\rm r}(t)] \right\} \nonumber \\
&= & \frac{\hbar}{2} ~ \int_{k_{\rm z} > 0} \frac{d^{3} {\bf
k}}{(2 \pi)^3} f^{\rm i}(k) \left [ {\bf R}({\bf k}) - {\bf
P}^{\rm i}\right ] \tau_k v_z~. \label{imprinting}
\end{eqnarray}
This is the spin density per unit area, so that the total spin
density per unit area in the semiconductor after reflection is
\begin{equation}
{\bf S} = n^{\rm i}~\frac{\hbar}{2}~{\bf P}^{\rm i} L + {\bf
S^{\rm r}}~, \label{volume}
\end{equation}
where $n^{\rm i} = \int \frac{d^{3}{\bf k}}{(2 \pi)^3} f^{\rm
i}(k)$ is the volume density of pumped electrons and $L$ is the
semiconductor length perpendicular to the interface. As shown in
Eq. (\ref{imprinting}), the contribution to the imprinted spin
from each wavevector channel is proportional to the mean-free path
$\tau_k v_z$.  Only the fraction of non-equilibrium electrons
within a mean-free path of the interface participate in the
spin-dependent reflection. This implies that for increasing sample
length $L$, the imprinted spin density per unit volume decreases
as $1/L$.  In the opposite limit, with the semiconductor length
$L$ shorter than the mean-free path, the situation is different
because multiple reflections occur and the electrons are quantum
confined. This was the impetus behind a recent proposal for a spin
valve with ferromagnetic gates \cite{ciuti2} and will be addressed
in a future publication.

After the non-equilibrium transient, the imprinted spin of the
semiconductor electron sea will decay with the long spin
relaxation time.  A weak magnetic field induces a Larmor
precession which is useful for measuring the imprinted spin. The evolution of
the spin in the semiconductor is governed by the Bloch equation
\begin{equation}
\frac{d{\bf S}}{dt} = \frac{g^\star \mu_{\rm B}}{\hbar} {\bf
B}_{\text {tot}} \times {\bf S}(t) - \frac{{\bf
S}(t)}{T_2^\star}~.
\end{equation}
The component of ${\bf S}$ orthogonal to the applied field ${\bf
B}$ can be extracted from the amplitude of the Larmor precession.
In addition, by measuring the effective Larmor frequency it is
possible to extract the nuclear field ${\bf B}_{\text{N}}$ (which
is proportional to the component of ${\bf S}$ along the magnetic
field).  There are three basic vectors in the problem that
determine the Faraday rotation: the external field ${\bf B}$, the
ferromagnet magnetization ${\bf M}$, and the injected pump
polarization ${\bf P}^{\rm i}$. Our theory can be tested by
systematically changing the relative orientations of these
vectors. We now discuss separately the cases of unpolarized and
polarized excitation.

\subsection{Unpolarized Excitation}

\begin{figure*}[tb]
\includegraphics[width=16cm]{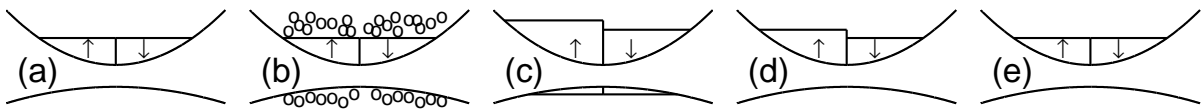}
\caption{The polarization process by  spin-reflection for
unpolarized excitation by a pump beam.  Explanation is in the
text. } \label{refl_cartoon}
\end{figure*}

The excited electron population is unpolarized, ${\bf P}^{\rm
i}=0$, when the non-equilibrium electrons are injected by a
linearly polarized laser or a non-magnetic electrical contact.
Although initially unpolarized, the non-equilibrium electrons will
be polarized by the reflection process.  This can be understood in
simple terms. Since the majority and minority spin electrons have
different wavevectors in the ferromagnet (i.e., the two spin bands
are exchange split), the reflection coefficients will in general
be different for the two spin channels.  This will leave a net
{\it spontaneous} polarization in the semiconductor.

The polarization process by spin-reflection for unpolarized
excitation from a pump beam is shown in Fig. \ref{refl_cartoon}
using a simplified model of a parabolic conduction band and a
single parabolic valence band.  (a) The semiconductor is lightly
$n$-doped to maximize the spin lifetime, so there is an
unpolarized background Fermi sea of electrons.  (b) A linearly
polarized pump beam excites spin-compensated non-equilibrium
electrons and holes.  The electrons reflect off the interface. The
different reflection coefficients of the two spin channels create
a net spin polarization in the semiconductor.  The holes undergo a
similar process, but the spin-orbit coupling rapidly relaxes any
polarization in the valence band.  (c) Energy and momentum
relaxation then drives the electrons and holes to the lowest
available states, with a net spin polarization remaining in the
conduction band.  (d) After the electron-hole recombination time,
all excess holes are gone, and a net polarization is left in the
Fermi sea of the conduction band.  (e) After the long
spin-relaxation time, the electron Fermi sea relaxes back to its
original unpolarized state.

The spin in the semiconductor after reflection will be
\begin{equation}
{\bf S}^{\rm r} = \frac{\hbar}{4} \hat{{\bf M}} \int_{k_z>0}
\frac{d^3{\bf k}}{(2\pi)^3} f^{\rm i}(k) (|r_{-,{\bf k}}|^2 -
|r_{+,{\bf k}}|^2) \tau_k v_z ,
\end{equation}
and hence is determined by the difference in the spin-dependent
reflectivities.  The spin asymmetry is quantum mechanical in
origin, so the sign of the polarization can be positive or
negative.  The coupling between the ferromagnet and semiconductor
is more efficient for either the majority or minority electrons
depending on the electronic structure of the ferromagnet and the
form of the interface potential in the semiconductor. The
spontaneous spin polarization is thus either parallel or
antiparallel to the ferromagnet magnetization ${{\bf M}}$.

Our theory offers a possible explanation of the findings of the
recent time-resolved Faraday experiments by Epstein {\it et
al.}.\cite{fmcoherence}  After pumping non-equilibrium unpolarized
electrons in the semiconductor, the experiment found that the
system quickly acquired (in tens of picoseconds, consistent with
the orbital relaxation time) a spontaneous spin polarization which
then decayed with the long spin relaxation time (nanoseconds). The
imprinted spin ${\bf S}^{\rm r}$ was aligned along ${\bf M}$ by
varying the angle between the applied field ${\bf B}$ and the
magnetization ${\bf M}$.  Interestingly, different ferromagnetic
materials gave different signs of the spin polarization relative
to ${\bf M}$.

\subsection{Polarized Excitation}

\begin{figure}
\includegraphics[width=8.8cm]{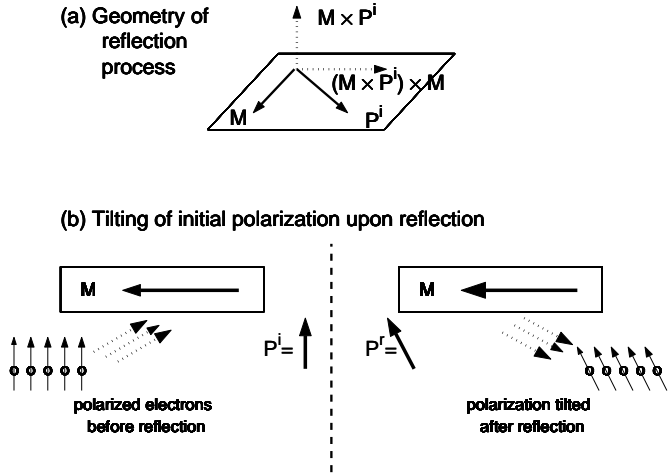}
\caption{(a) The geometry of the reflection process for polarized
excitation.   In general, the reflected polarization has
components along all three spatial directions.  (b) A net spin
polarization present before reflection will be tilted upon the
interaction at the interface. } \label{tilt_cartoon}
\end{figure}

When the electrons are pumped with a net polarization ${\bf
P}^{\rm i} \ne 0$, the reflected spin does not align with ${\bf
M}$ or ${\bf P}^{\rm i}$. There is an extra component which is
perpendicular both to ${\bf M}$ and ${\bf P}^{\rm i}$, i.e., the
reflection off the ferromagnet produces a spin-torque. The
geometry of the process is depicted in Fig. \ref{tilt_cartoon}(a).
The amplitude of the extra component is proportional to ${\rm
Im}(r_- r_+^*)$, implying that the spin torque is due to the
different reflection phase-shift for minority and majority spin
channels. A pre-existing spin polarization vector can be
manipulated using this property of the reflection process, as is
shown schematically in Fig. \ref{tilt_cartoon}(b).

In principle, this kind of effect can be detected through
time-resolved Faraday rotation experiments with circularly
polarized light.   Tilting of the optically injected polarization
vector due to reflection would appear as a phase-shift in the
Larmor precession. The spin torque term, which is aligned along
${\bf M} \times {\bf P}^{\rm i}$, would change sign if the
magnetization were switched, ${\bf M} \rightarrow -{\bf M}$,
resulting in a jump in the phase of the Larmor precession.
Experimental studies of this phase shift have been observed.
However, a systematic study as a function of the relative
geometric orientation of optically-injected polarization,
ferromagnet magnetization and magnetic field are necessary to
fully test our theoretical predictions.

The {\it spin torque} term we describe is analogous to an effect
predicted by Slonczewski \cite{slon2} and recently demonstrated in
all-metallic systems.\cite{torque}  A spin-polarized current
injected into a ferromagnet can tilt the magnetization of the
ferromagnet, provided that (1) the ferromagnet magnetization and
the polarization of the current are not collinear, and (2) the
spin-polarized current injected into the ferromagnet is quite
large. However, in the system that we consider, the density of
carriers in the semiconductor in so small compared to the density
of carriers in the ferromagnet that the tilting of the ferromagnet
magnetization is negligible.  Instead we find complementary
behavior in which the ferromagnet magnetization tilts the
polarization of the current.

\section{Results for a Schottky barrier}
\label{example}

In this section we investigate in detail the spin reflection at
the Schottky barrier between a homogeneously $n$-doped
semiconductor and a ferromagnetic metal. We define the zero of the
potential energy at the Fermi level in the metal. Within the
depletion layer approximation, the potential energy in the
semiconductor space-charge region ($-z_{\rm b} < z < 0$) is given
by
\begin{equation} \label{eq-sp}
U(x) = V- E^{\rm {sc}}_{\rm {f}} + (U_{\rm b}+ E^{\rm {sc}}_{\rm
{f}}-V) \left ( 1+\frac{z}{z_{\rm b}} \right )^2 ~,
\end{equation}
where $V$ is the applied bias, $E^{\rm {sc}}_{\rm {f}}$ is the
Fermi kinetic energy in the semiconductor, $U_{\rm {b}}$ is the
Schottky barrier height, and $z_{\rm {b}} = \sqrt{\epsilon_0
U_{\rm {b}}/(2 \pi n e^2)}$ is the depletion width. For high bias
$ V
> U_{\rm {b}} + E^{\rm {sc}}_{\rm {f}}$, the semiconductor is
considered at flat band. The depletion approximation could be
replaced with a more realistic model, such as one constructed
using the coupled Poisson and Thomas-Fermi equations, but we have
explicitly calculated and verified that the deviation from the
depletion approximation is negligible.

In Ref.~\onlinecite{ciuti1} we gave an analytical approximation
for the spin-dependent reflection using an effective rectangular
barrier. In this paper, we show the exact numerical results for
the realistic Schottky potential in Eq.~(\ref{eq-sp}). We also
present results for the optical excitation of non-equilibrium
electrons, in which pumping takes place in the barrier region, and
contrast them with the results for excitation at the Fermi level
(electrical injection). These are of interest since both transport
and optical experiments have been performed on Schottky barriers
in which spin reflection plays a role.

\subsection{Unbiased Schottky barrier}

\begin{figure}[t!]
\includegraphics[width=8cm]{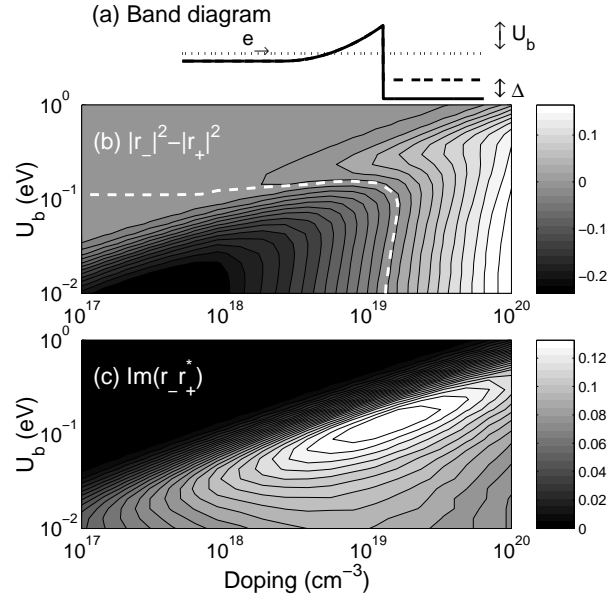}
\caption{Reflection from the Schottky junction in equilibrium. (a)
The band diagram for the calculation.  The electrons are incident
from the Fermi level in the semiconductor.  (b) Contours of the
spin reflection asymmetry $|r_-|^2-|r_+|^2$ as a function of the
semiconductor doping $n$ and the Schottky barrier height $U_{\rm
b}$.  This quantity changes sign at the white dashed line.  (c)
The component of reflected polarization orthogonal to both $\bf M$
and ${\bf P}^{\rm i}$, which is always positive. }
\label{schottky_prl}
\end{figure}

Under no applied bias, $V=0$, the semiconductor and the
ferromagnet are in equilibrium, so that the Fermi levels in the
two systems are equal.  We consider the spin-dependent reflection
of electrons which are incident on the Schottky barrier from the
semiconductor side with the Fermi kinetic energy $E_{\rm f }^{\rm
{sc}} = \hbar^2 (3 \pi^2 n)^{2/3} / (2 m^\star_{\rm {sc}})$. The
spin-dependence of the reflection coefficient arises from the
spin-dependent Fermi velocities in the ferromagnet, which are, in
the two-band model, $v^{\text{fm}}_{+} = \sqrt{2
E^{\text{fm}}_{\text{f}}/m^{\star}_{\text{fm}}}$ and
$v^{\text{fm}}_{-}= \sqrt{2
(E^{\text{fm}}_{\text{f}}-\Delta)/m^{\star}_{\text{fm}}}$ for the
majority and minority spin, respectively.
$E^{\text{fm}}_{\text{f}}$ and $m^{\star}_{\text{fm}}$ are the
Fermi energy and effective mass in the ferromagnet.

The numerical results for the reflection coefficients have been
obtained through a finite-difference solution of the Schr\"odinger
equation.\cite{liu} The input parameters  for the semiconductor
are those of bulk GaAs ($m^\star_{\rm {sc}}=0.07~ m_0$, $\epsilon
= 12.9$).  For the ferromagnet, we use the values  $m^\star_{\rm
{fm}}=1$, $E_{\rm f}^{\rm {fm}}=2.5~ e{\rm {V}}$, $\Delta=1.9~
e{\rm {V}}$).  We show the results for the spin reflection
difference $|r_-|^2-|r_+|^2$ and the spin torque amplitude $\rm
{Im}(r^\star_- r_+)$ as  functions of the semiconductor doping $n$
and Schottky barrier height $U_{\rm {b}}$ are shown in
Fig.~\ref{schottky_prl}.

Fig.~\ref{schottky_prl}(b) shows the spin difference
$|r_-|^2-|r_+|^2$, which  is the spin polarization generated by
reflection, directed along the ferromagnet magnetization, ${\bf
M}$. Both the Schottky barrier height and the semiconductor doping
have significant impact on this generated spin polarization.
Variation of either property within experimentally accessible
values  can cause a change of direction of the spin polarization
(indicated by the white-dashed line). The numerical results for
the realistic Schottky potential show that our approximated
solutions with an effective rectangular barrier in
Ref.~\onlinecite{ciuti1} are very close. The shape of the contours
are identical and only minor discrepancies are present. Note that,
even in the absence of a barrier, the velocity mismatch at the
interface leads to spin-dependent reflection.  At low doping
$n<10^{19} \text{cm}^{-3}$, the semiconductor velocity is better
matched to the minority spin band velocity, but at high doping
$n>10^{19} \text{cm}^{-3}$, the semiconductor velocity is better
matched to the majority spin band velocity (since the Fermi energy
for minority spins is smaller than the Fermi energy for majority
spins).  This is the reason for a change of sign in the spin
difference as a function of doping.  This argument may be extended
to interfaces with small Schottky barriers, but for large Schottky
barriers tunneling plays a dominant role.
Fig.~\ref{schottky_prl}(b) shows the sign change of the spin
difference when the barrier height is larger than about $0.1~
e\text{V}$ that joins smoothly with the change of sign  from the
low-barrier region.  For large Schottky barriers, the large
wave-vector in the barrier region better matches the majority spin
band velocity, inducing the change of sign.

 Figure ~\ref{schottky_prl}(c) shows the spin torque term for
polarized incident electrons. This is the reflected polarization
component orthogonal to both the ferromagnet magnetization ${\bf
M}$ and the electron polarization ${\bf P}^{\rm i}$.  We find that
this term always has the same sign and is analogous to the
Kramers-Kronig partner of the spin difference shown in
Fig.~\ref{schottky_prl}(b).

\begin{figure}[t!]
\includegraphics[width=8cm]{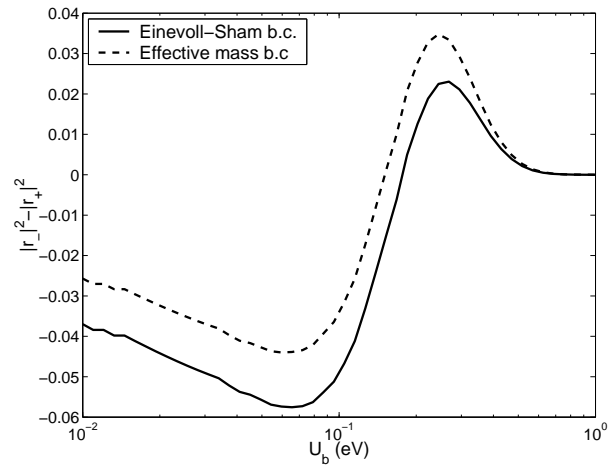}
\caption{A comparison of the spin reflection asymmetry using the
effective mass boundary condition and the Einevoll-Sham
\cite{einevoll} boundary condition.  The semiconductor density is
fixed at $n=10^{19}~\rm{cm}^{-3}$.} \label{einevoll}
\end{figure}

In the effective mass model used above for both the metal and the
semiconductor, we match the wave function and its derivative with
the only accommodation of the effect the different crystals  being
the effective masses($\psi^\prime / m^\star$ is continuous). We
have investigated the effect of the semiconductor band structure
effect in the limit of small gap with the boundary condition
\cite{einevoll} which modifies the envelope wave function slope on
the semiconductor side by a factor of a half,
\begin{equation}
\frac{1}{2m^*_{\text{sc}}} \frac{\psi'_{\text{sc}}}{\psi_{\text{sc}}}
= \frac{1}{m^*_{\text{fm}}} \frac{\psi'_{\text{fm}}}{\psi_{\text{fm}}}.
\end{equation}
We have found that the spin property dependence on the barrier
height and doping density qualitatively unchanged (see Fig.
\ref{einevoll}).

Our theory implies that the reflected spin polarization is very
dependent on the band profile in the semiconductor.  The Schottky
barrier height $U_{\rm b}$ and barrier width as measured by the
semiconductor doping $n$ have a large impact on both the sign and
magnitude of the polarization.  Both effects can be used to {\it
tailor} the Schottky barrier to achieve the desired spin
polarization through different ferromagnetic materials and
different doping concentrations. Our results also show that spin
polarization generation through the transmission or reflection of
a barrier is a quantum mechanical phenomenon.  As is evident from
our plot, the reflection can not simply be modelled as depending
on the ferromagnetic density of states, or as a spin-dependent
resistance that is insensitive to the exact properties of the
barrier.

\subsection{The effect of an applied bias}
\begin{figure}[t!]
\includegraphics[width=8cm]{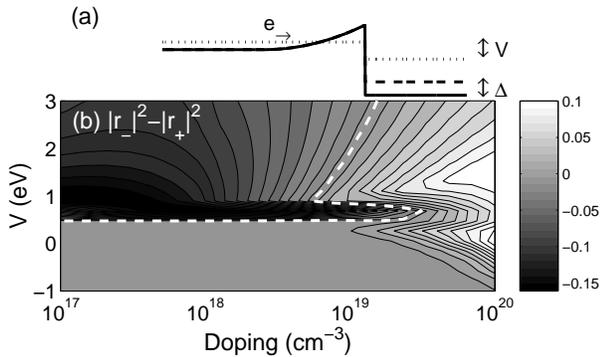}
\caption{ Reflection from the biased Schottky barrier.  (a) The
band diagram.  For negative bias, the semiconductor Fermi level is
below the ferromagnet Fermi level.  For $V > (U_{\rm b}+E_{\rm
f}^{\rm {sc}})$ the semiconductor is in flat band condition and
the electrons see no barrier.  The Schottky barrier height is
fixed at $U_{\rm b} = 0.7~e\rm{V}$.  (b)  The spin reflection
asymmetry as a function of semiconductor doping $n$ and applied
bias $V$.  The change of sign is indicated by the dashed white
line, and its horizontal segment is near the Schottky barrier
height. } \label{schottky_bias}
\end{figure}

Now we examine the effect of applying a bias across the Schottky
barrier.  We assume that applying a positive bias $V>0$ raises the
semiconductor Fermi level above the ferromagnet Fermi level by an
amount $V$, i.e., the entire voltage drop is across the barrier;
the semiconductor bulk and the ferromagnet bulk are assumed to be
flat.   The bias may be viewed approximately as just changing the
Schottky barrier height, so that for negative bias the Schottky
barrier is larger by $V$ and for positive bias the Schottky
barrier is smaller by $V$.  For $V>(U_{\rm b}+E_{\rm f}^{\rm
{sc}})$, the barrier disappears and the entire semiconductor
region is a flat band. Moreover, since the tunnelling is a
ballistic process, the bias implies that the transmission into the
ferromagnet is not at the Fermi level in the ferromagnet.  Thus
the applied bias changes the majority and minority velocities and
hence the matching of the semiconductor electrons (at the Fermi
level in the semiconductor $E_{\rm f}^{\rm {sc}}$). In this
calculation we choose the Schottky barrier height to be $U_{\rm b}
= 0.7~e\rm{V}$, the value of Fe/GaAs.

In Fig.~\ref{schottky_bias} we plot the spin reflection difference
$|r_-|^2-|r_+|^2$ for unpolarized excitation.  At $V=0$, this
corresponds to a slice out of Fig. \ref{schottky_prl} at the
Schottky barrier height $U_{\rm b}=0.7~e\rm{V}$.  As  the figure
shows, changing the bias  makes a change of sign in the spin
reflection difference in a similar way to the change of sign as a
function of Schottky barrier height in Fig.~\ref{schottky_prl}(b).
For negative bias $V<0$, the Schottky barrier increases in both
height and width. The net effect is that the reflection
coefficients for both spin channels are very nearly unity, so that
the difference between them is negligible. The sign of the
polarization is the same as the sign at zero bias, but the spin
difference goes asymptotically to zero. The sign of the
polarization changes abruptly for positive bias around $V \approx
.5~e\rm{V}$, then rises rapidly to its maximum value. This is
because the bias makes the Schottky barrier smaller until it
disappears completely;  the potential becomes a step and the
velocity mismatch becomes the important factor. Hence, the high
positive bias regime is similar to the low barrier regime in
Fig.~\ref{schottky_prl}(b). As the bias increases further, the
reflection difference decreases, due to the concomitant increase
in electron energy in the ferromagnet, so that the exchange
splitting $\Delta$ becomes less effective.

These results suggest that a bias could be used to control the
sign and magnitude of the spin polarization.  For a fixed doping,
the reflected spin polarization can be tuned depending on the
applied bias.  However, it is much easier to get high polarization
for one sign of polarization than the other (in
Fig.~\ref{schottky_bias}(b), the polarization is maximum above the
change of sign, but is very nearly zero below the change of sign
for low doping).

\subsection{Special effects of optical excitation}
\begin{figure}[t!]
\includegraphics[width=8cm]{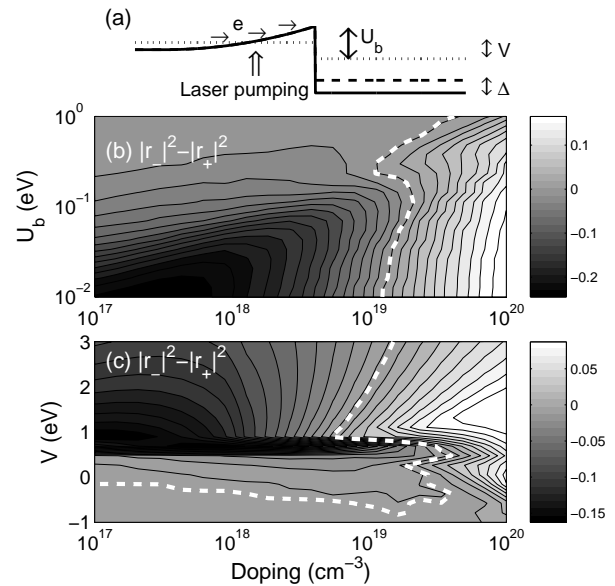}
\caption{ Reflection from the Schottky barrier under optical
excitation.  (a) The band diagram.  Electrons are taken to be
excited homogeneously in space, including in the depletion region.
(b) The spin reflection asymmetry as a function of semiconductor
doping $n$ and Schottky barrier height $U_{\rm b}$, keeping the
applied bias at $V=0$.  (c) The spin reflection asymmetry as a
function of semiconductor doping $n$ and applied bias $V$, keeping
the Schottky barrier height fixed at $U_{\rm b} = 0.7~e\rm{V}$. }
\label{schottky_optical}
\end{figure}

Optical pumping creates special effects in spin reflection at a
Schottky barrier not seen under electrical injection. In the
latter case, non-equilibrium electrons are injected close to the
Fermi level far from the barrier whereas optical excitation
creates electrons everywhere in the semiconductor, including in
the space charge region. This has a profound effect on the net
spin polarization because not all electrons will see the same
effective barrier height.  Electrons very near the interface will
see a much smaller Schottky barrier than electrons pumped far away
from the interface.  Since the electrons closest to the interface
see a smaller and narrower barrier, they will couple to the
ferromagnet much more efficiently, so that the polarization is
dominated by these electrons. This effect is quantitatively very
important because of the exponential dependence of the tunnelling
coupling on the barrier height and width.  The results of optical
excitation are averaged over the different barrier heights.

To fully grasp the difference between optical and electrical
excitation, we attempt to quantify the optical effect. First, we
assume that the electrons are pumped homogeneously in the space
charge region (no contribution from the bulk region is included).
In the experiment,\cite{fmcoherence} the semiconductor is thin
enough for this to be approximately correct.  The laser bandwidth
is taken to be approximately $10~{\rm m}e{\rm V}$ (the typical
bandwidth of a pulsed laser in time-resolved pump-probe
experiments). We calculate the reflection for a Gaussian
distribution of electron energies for each coarse-grained spatial
neighborhood in the space charge region. We then average these
effects over the different barrier heights due to the homogeneous
pumping along the space charge region (see
Fig.~\ref{schottky_optical}(a)).

Results for the spin refection difference $|r_-|^2-|r_+|^2$ from unpolarized
excitation are shown in Fig.~\ref{schottky_optical}(b) as a
function of doping and Schottky barrier height keeping the bias at
$V=0$. Compare the plot with the case of excitation at the
Fermi level, Fig.~\ref{schottky_prl}(b).
 There is not much change for the low barrier region
because the electrons pumped high in the barrier do not see a
drastically different barrier from the electrons pumped farther
away. On the other hand, at high barrier heights, there are drastic
changes. The
change of sign of the polarization disappears for low doping
concentrations. In the
case of electrical excitation, at a high barrier all the electrons see a
very high
and thick barrier which allows for very little tunnelling. In the case of
optical pumping, there are some electrons pumped high in the
barrier which see a very small barrier and are efficiently coupled
to the ferromagnet.  The polarization at low barrier is of
opposite sign to the polarization at high barrier, but, since the
coupling for low barrier is so much more efficient, it dominates
and the change of sign disappears.

Figure~\ref{schottky_optical}(c)  shows the spin reflection difference at a
constant
 Schottky barrier height,  $U_{\rm{b}}= 0.7~e{\rm{V}}$,
and varying doping and applied bias.  In comparison
with Fig.~\ref{schottky_bias}(b),  the optical excitation
drastically changes the qualitative aspects of the spin reflection
difference. For low doping, the change of sign as a function of
bias has been pushed down to negative bias for the same
reason as the disappearance of the change of sign in
Fig.~\ref{schottky_optical}(b)
for low doping.  The spin reflection difference is dominated by the
electrons pumped
high in the barrier, which have the opposite sign of polarization compared with
the electrons near the Fermi level in the semiconductor.

There is a caveat to these special effects.  A more realistic
model would use an electron wavepacket to reflect off the
interface in order to  account for the significant scattering due
to efficient optical phonon emission. The electrons pumped high in
the barrier would be in a region with almost no background
electron density, so that the spin lifetime may be shorter than
the lifetime outside the depletion region. Our purpose here is to
point out that electrical and optical experiments can give
inconsistent results if it is assumed that the underlying
mechanism is exactly the same in both cases.  Electrically excited
electrons stay close to the Fermi level, while optically excited
electrons exist all over the barrier.  The pumping in the barrier
drastically alters the spin reflection difference which would be
observed in in experiments on the interface, and hence care must
be taken when comparing optically excitation experiments with
electrical injection experiments. In particular, the optically
excited electrons pumped high in the barrier can have different
sign of polarization to electrons lower in the barrier, making it
difficult to compare polarizations from experiments involving
optical versus electrical injection.

\section{Conclusions}
\label{conclusions}

We have examined the electronic properties
associated with semiconductor electrons that are in proximity with
a ferromagnet.  Instead of focusing on the injection of
ferromagnet electrons into the semiconductor through the
interface, we have instead chosen to look at the properties of
semiconductor electrons that can interact with a ferromagnetic
epilayer.  We have focused on  the reflection of
non-equilibrium electrons from the interface.

For doped semiconductors with ferromagnetic epilayers, we have
calculated the spin-dependent reflectivities for electrons
incident on the Schottky barrier from the semiconductor.  We find
that the shape of the barrier has a large impact on the asymmetry
between reflection for the two spin channels, and significantly
that the sign of the difference can change depending on the system
parameters.  We have included calculations that mimic the behavior
of optical excitation, which can radically affect the spin
reflection difference.  The ability to control the sign and
magnitude of the spin polarization by tuning the properties of the
Schottky barrier (through the semiconductor doping or through an
applied bias, for example) may be useful for device design.

In contrast to spin injection from a ferromagnet into a
semiconductor, spin reflection has the advantage that the
processes are kept in the semiconductor.  For carriers confined
near the interface, multiple reflections from the interface can
enhance the single-reflection spin asymmetry, yielding larger
polarizations.   Such reasoning has led us to propose a spin-valve
device with ferromagnetic gates.\cite{ciuti2} A detailed study of
the coupling of the equilibrium and transport properties of a
confined semiconductor electron system in contact with the
ferromagnet will be the subject of another long paper.

\begin{acknowledgments}
This work is supported by DARPA/ONR N0014-99-1-1096,
NSF DMR 0099572, the Swiss National Foundation (for CC), and University
of California Campus-Laboratories Cooperation project
 (for JPM).  We thank
D.D. Awschalom, R.J. Epstein and R. Kawakami for helpful discussions.
\end{acknowledgments}

\end{document}